\documentclass{aa}
\usepackage{graphicx}
\begin{document}

\title{On the evolutionary status of chemically peculiar stars
of the upper main sequence\thanks{Based on observations by the Hipparcos
satellite}}
\author{H.~P\"ohnl\inst{1}, H.M.~Maitzen\inst{1}, E.~Paunzen\inst{1,2}}

\mail{maitzen@astro.univie.ac.at}

\institute{Institut f\"ur Astronomie der Universit\"at Wien,
           T\"urkenschanzstr. 17, A-1180 Wien, Austria
\and       Zentraler Informatikdienst der Universit\"at Wien,
           Universit\"atsstr. 7, A-1010 Wien, Austria}

\date{Received 19 November 2002/Accepted 28 January 2003}
\titlerunning{On the evolutionary status of chemically peculiar stars}{}

\abstract{We present further evidence that the magnetic chemically peculiar
stars (CP2) of the upper main sequence already occur at very early stages of the
stellar evolution, significantly before they reach 30\,\% of their life-time
on the main sequence. This result is especially important for models
dealing with dynamo theories, angular momentum loss during the pre- as well
as main sequence and evolutionary calculations for CP2 stars. Results from the
literature either derived for objects in the Hyades and the UMa cluster or
from the Hipparcos mission contradict each other. A way out of this dilemma is
to investigate young open clusters with known ages and accurate distances (error\,$<$\,10\%),
including CP2 members. Up to now, four open clusters fulfill these
requirements: IC~2391, IC~2602, NGC~2451\,A and NGC~2516. In total, 13 CP2
stars can be found within these clusters. We have used the measurements and
calibrations of the Geneva 7-color photometric system to derive effective temperatures
and luminosities. Taking into account the overall metallicity of the individual
clusters, isochrones and evolutionary tracks were used to estimate ages and masses
for the individual objects. The derived ages (between 10 and 140\,Myr) are well in line
with those of the corresponding clusters and further strengthen the membership of
the investigated CP2 stars.
}
\maketitle

\section{Introduction}

A century ago, Antonia Maury (1897) detected a subclass of A-type stars with
peculiar lines and line strengths which thereafter became known as Ap (or CP)
stars. Subsequently, these stars revealed other peculiar features,
e.g. the existence of a strong global magnetic field (Babcock 1947) with a predominant
dipole component located at random with respect to the stellar rotation
axis and the center of the star as well as overabundances in respect to the Sun
for heavy elements such as silicon, chromium, strontium and europium. 
This lead to
the Oblique Rotator concept of slowly rotating stars with non-coincidence
of the magnetic and rotational axes (Stibbs 1950). 

The peculiar (surface) abundances for CP stars have been explained
by the following two theories: 1)
{\it Diffusion of chemical elements} depending on the balance between
gravitational pull and uplift by the radiation field through absorption
in spectral lines (Michaud 1970). 
2) {\it Selective Accretion} from the interstellar medium (Havnes \&
Conti 1971) via the stellar magnetic field.

The origin of the magnetic fields is still a matter of debate: those who
favor the survival of frozen-in fossil fields originating from the
medium out of which the stars were formed are in opposition to those
following the idea that a dynamo mechanism is acting in the interior of
these stars. 

For many decades the evolutionary status of the CP stars has been
controversial. Oetken (1984) concluded that the CP2 (magnetic
CP stars) phenomenon
appears at the late stages of the main sequence evolution. This theory
was further strengthened by the results of Hubrig \& Schwan (1991) and Hubrig
\& Mathys (1994) who analysed peculiar objects in the Hyades and the UMa cluster.
Hubrig et al. (2000) found that the distribution of CP2 stars of masses below 
3\,M$_{\odot}$ in the Hertzsprung-Russell-diagram differs from that of the ``normal'' 
stars in the same temperature range at a high level of significance: magnetic stars are 
concentrated toward the center of the main sequence band. In particular, they found that 
magnetic fields appear only in stars which have already completed at least 
approximately 30\% of their main sequence life-time. No clear picture emerges as to the 
possible evolution of the magnetic field across the main sequence. Hints of some (loose) 
relations between magnetic field strength and other stellar parameters are found: stars 
with shorter periods tend to have stronger fields, as do higher temperature and 
higher mass stars. A marginal trend of the magnetic flux to be lower in more slowly 
rotating stars may possibly be seen as suggesting a dynamo origin for the magnetic
field. No 
correlation between the rotation period and the fraction of the main sequence life time 
completed is observed, indicating that the slow rotation in these stars must already 
have been achieved before they became observably magnetic.

The results of the Hipparcos mission on the other hand do not support the 
mentioned above findings. G\'omez et al. (1998)
presented the Hertzsprung-Russell-diagram
of about 1000 CP stars in the solar neighbourhood 
using astrometric data from Hipparcos satellite as well as 
photometric and radial velocity data. Most CP stars lie on the main sequence 
occupying the whole width of it (about 2 mag), just like ``normal'' stars in the 
same range of spectral types. Their kinematic behaviour is typical of thin 
disk stars younger than about 1 Gyr. 

North (1993), North et al. (1997) and Wade (1997) reported that
the CP stars are distributed uniformly along the width of the main sequence with
strong magnetic fields existing at all evolutionary states.
Stepien (1998) found no evidence that CP2 stars
undergo significant angular momentum loss during their main sequence lifetime.

We have investigated CP2 stars of young open clusters in order to tackle the
question of whether young peculiar objects exist or not and to put further constraints
on theoretical models. Members of open clusters allow to determine precise luminosities
and effective temperatures based on reliable calibrations. With the knowledge of 
accurate distances (e.g. from Hipparcos measurements) and estimates of the overall
metallicities, corresponding isochrones can be fitted, resulting in masses and ages
for individual CP2 objects.

We have chosen young open clusters (age\,$<$\,100\,Myr) including at least one
CP2 with accurate proper motions and distances (error less than 10\,\%) for our 
investigation. In total, four open clusters (IC~2391, IC~2602, NGC~2451\,A and NGC~2516)
fulfill our requirements.

\begin{figure}
\begin{center}
\includegraphics[width=76mm]{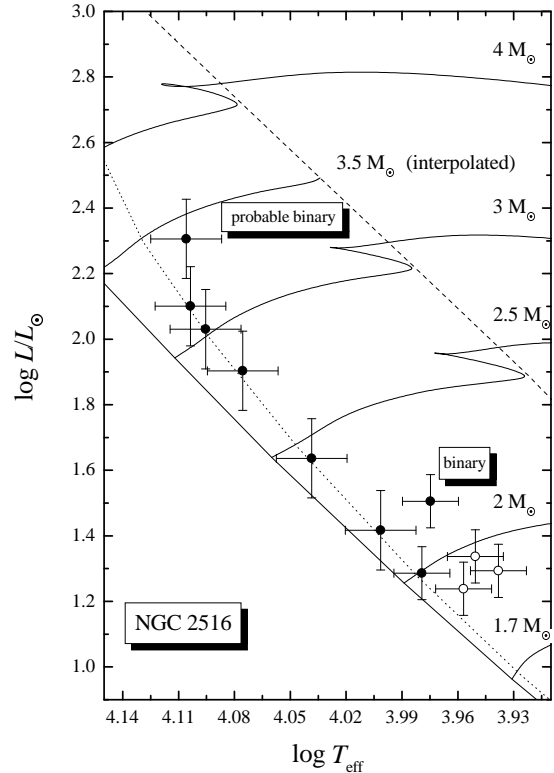}
\caption{The location of the CP2 stars for NGC~2516 taken from Table\,\ref{cp2_res}.
Filled circles are well established CP2 objects and open circles are doubtful
cases (CP$-$60\,948, CP$-$60\,1008 and CP$-$60\,1013).
The dotted line is the isochrone for log\,$t$\,=\,8.00 and $[$Fe/H$]$\,=\,$-$0.23
taken from Schaller et al. (1992) whereas the dashed line denoted the terminal age
main sequence; the evolutionary tracks for individual masses are interpolated
within the ones listed by Schaller et al. (1992), Schaerer et al. (1993a,b) and 
Charbonnel et al. (1993).}
\label{ngc2516}
\end{center}
\end{figure}

\begin{table*}[t]
\begin{center}
\caption{Data about our program clusters. The errors in the final digits of the 
corresponding quantity are given in parenthesis.}
\label{all_res}
\begin{tabular}{lcccc}
\hline
\hline
Name & NGC~2451~A & NGC~2516 & IC~2391 & IC~2602  \\
     & C0743$-$378 & C0757$-$607 & C0838$-$528 & C1041$-$641 \\
\hline
$l/b$ & 252/$-$7 & 274/$-$16 & 270/$-$7 & 290/$-$5 \\
$E(B-V)$ & 0.01 & 0.13 & 0.01 & 0.04 \\
$\pi$(Lit) & 5.31(19) & 2.89(21) & 6.85(22) & 6.58(16) \\
$\pi$(our) & 5.35(21) & 2.69(22) & 6.98(18) & 6.99(15) \\ 
$d$\,[pc] & 187(7) & 372(33) & 143(3) & 143(3) \\
log\,$t$ & 7.70 & 8.00 & 7.70 & 7.46 \\
$[$Fe/H$]$ & $-$0.26 & $-$0.23 & $-$0.04 & $-$0.20 \\
n(CP2) & 1 & 8 & 2 & 2 \\
\hline
\end{tabular}
\end{center}
\end{table*}

\section{Program clusters}

A main criterion for our target selection was the knowledge of accurate
distances. Robichon et al. (1999) have presented parallaxes with errors less
than 10\,\% for nine young open clusters (age\,$<$\,100\,Myr) of which
four have known CP2 members. We have not investigated older open clusters 
because members of such aggregates have evolved more than 30\,\% of their main
sequence life-time. Such CP2 stars are already known in the Galactic field.

We have slightly modified the values listed in Robichon et al. (1999). They
have taken all objects within 10\,pc of the individual cluster centers for
their determination of the distances. For our work, only objects within three times
the cluster radius (taken from Lyng\aa\,\,1987) were considered. Furthermore,
we have discarded objects with proper motions deviating more than 
0.221\,$\pi$\,[mas\,yr$^{-1}$] (=\,1\,kms$^{-1}$) from the calculated overall 
proper motion of the individual open
cluster. Table\,\ref{all_res} list these parallaxes with those of Robichon et al. (1999)
and other characteristics of our program clusters. All parallaxes agree very well
and are also consistent with results from van Leeuwen (1999).

\subsection{IC~2391}

This young open cluster contains two CP2 stars according to North (1984)
and Maitzen \& Catalano (1986). Both objects are members of IC~2391
(Perry \& Hill 1969; numbering system).
\begin{itemize}
\item HD~74169 (\#18): classified as Ap\,EuCr(Sr) by Buscombe (1965), Houk (1978)
and Levato \& Malaroda (1984).  
\item HD~74535 (\#31): peculiar according to the same references as listed
above.
\end{itemize}
The values for the metallicity and reddening were taken from Lyng\aa\,\,(1987).

\subsection{IC~2602}

The overall characteristics of this open cluster was only investigated by 
Lyng\aa\,\,(1987) who lists log\,$t$\,=\,7.46, $E(B-V)$\,=\,0.04 and 
[Fe/H]\,=\,$-$0.20. The numbering in brackets is according
to Braes (1962).
\begin{itemize}
\item HD~92385 (\#71): Bidelman \& McConnell (1973) and Houk \& Cowley (1975)
establish the membership and the peculiar nature which was confirmed by
Maitzen et al. (1988).
\item HD~92664 (\#27): additional classification as B9p\,(Si) by Whiteoak (1961)
and Abt \& Morgan (1972).
\end{itemize}
The $\Delta a$ measurements of Maitzen et al. (1988) for both objects show 
only moderate peculiarity.

\subsection{NGC~2451\,A}

What was believed to be one open cluster was actually separated into one
less and one more distant aggregate within one line of sight (Eggen 1983,
Maitzen \& Catalano 1986). R\"oser \& Bastian (1994) postulated two clusters
at distances of 220 and 400\,pc which was later confirmed by Platais et al. (1996).
Carrier et al. (1999) used the Hipparcos data and found two similar aged
clusters (log\,$t$\,=\,7.70) at distances of 197 (NGC~2451\,A) and 358\,pc (NGC~2451\,B).
For NGC~2451\,A they list 28 members including the peculiar object
HD~63401 (\#277; Williams 1967). This object was classified as Ap\,Si by Bidelman
\& McConnell (1973) and Houk (1982) whereas Hartoog (1976) lists B8III. 
The photometric CP2 star HD~62992 (\#250) is a non-member according to
Carrier et al. (1999).

\subsection{NGC~2516}

This open cluster contains at least eight CP2 stars for which spectral 
classification resolution spectroscopy as well as $\Delta a$ photometry is
available (Maitzen \& Hensberge 1981). Mermilliod (1981) gives an age 
log\,$t$\,=\,8.03 whereas Lyng\aa\,\,(1987) lists a value of 7.85. 
Jeffries et al. (1997) derived a metallicity of [Fe/H]\,=\,$-$0.32(6)
for cooler members. Additional estimates were made by Cameron (1985;
$-$0.42), Lyng\aa\,\,(1987; $-$0.23) and Debernardi \& North (2001; 0).
The metallicity calibration of Eggen (1972) for the Johnson $UBV$ system
results in a value of $-$0.23 whereas Str\"omgren $uvby$ photometry gives
$-$0.20 (Snowden 1975, Berthet 1990). Taking all these values into account,
we have chosen to use the value of Lyng\aa\,\,(1987) as a ``mean''
metallicity: [Fe/H]\,=\,$-$0.23.

Maitzen \& Hensberge (1981) have reported a significant differential reddening
throughout the cluster area. We have used the $Q$-method (Johnson \& Morgan 1953)
to calculate the overall reddening as well as the individual values for
the CP2 stars. The derived values are well in agreement with the results listed
by Lyng\aa\,\,(1987).

The following stars are probably CP2 stars (the numbering is according to
Cox 1955):
\begin{itemize}
\item HD~65712 (not listed in Cox 1955): Dachs \& Kabus (1989) list a spectral type of
A0p\,(Si\ion{II}) and a high membership probability consistent with the results of
Bidelman \& McConnell (1973). Maitzen \& Hensberge (1981) measured a very high $\Delta a$-value
of +70\,mmag.
\item HD~65987 (\#15): A possible binary system (Abt \& Levy 1972, Snowden 1975,
Gieseking 1978) classified as B9.5IVp\,(Si\ion{II}) and B9p\,Si by Abt \& Morgan (1969)
and Hartoog (1976), respectively. 
\item HD~66295 (\#26): B8/9IV\,Si according to Houk \& Cowley (1975), listed as member
in Dachs \& Kabus (1989).
\item HD~66318 (\#24): member, classified as A0p\,EuCrSr (Hartoog 1976).
\item CP$-$60\,944A (\#208) and B (\#209): the objects have spectral types of
B8\,III and B8p\,Si with an apparent confusion about the identification of the
individual components.
\item CP$-$60\,978 (\#c): a member (Dachs \& Kabus 1989) classified as A0p\,(Si\ion{II})
and A0p\,EuCrSr (Abt \& Morgan 1969; Hartoog 1976).
\item CP$-$60\,981 (\#38): visual binary with a period of 3.2\,d (North 1984) with a
classification of A2Vp\,(SrCrEu) given by Hartoog (1976).
\item CP$-$60\,948 (\#8), CP$-$60\,1008 (\#17) and CP$-$60\,1013(\#28): 
Maitzen \& Hensberge (1981) measured $\Delta a$ of +27 (\#17) and +13\,mmag
(\#8 and \#28); no other data in the literature were found.
\end{itemize}
The apparent peculiarity of CP$-$60\,948 (\#8), CP$-$60\,1008 (\#17) and CP$-$60\,1013(\#28)
should be confirmed by further spectroscopic data.

New observations at the VLT (Wade, private communication) revealed a very strong
magnetic field (4.5\,kG longitudinal field, 15\,kG surface field) in HD 66318. Weaker
fields were found for HD 66295 ($-$0.8\,kG longitudinal) and CP$-$60\,944A (0.6\,kG). No
magnetic fields were detected for CP$-$60\,948, CP$-$60\,981 or any other above listed
objects. These results are in excellent agreement with the measurements within the
photometric $\Delta a$-system.

\begin{table*}[t]
\begin{center}
\caption{Summary of results for true CP2 stars in our program clusters; 
HD~65987 and CP$-$60\,981 are binary systems for
which we list the values for the primary component. 
The errors in the final digits of the corresponding quantity are given in parenthesis;
$\tau$ is the age of the star, $\tau_{HR}$ the time for a star on the main sequence and
$\tau_{Cl}$ the known age of the individual cluster.}
\label{cp2_res}
\begin{tabular}{lcccccccc}
\hline
\hline
\vspace{-3mm}
\\
Cluster & $[$Fe/H$]$ & Object & log\,$T_{\rm eff}$ & log\,$L/L_{\sun}$ & $M/M_{\sun}$ & 
$\tau$ & $\tau_{HR}$ & $\tau_{Cl}/\tau_{HR}$ \\
& [dex] & & & & & [Myr] & [Myr] \\
\hline
NGC~2451\,A & $-$0.26 & HD~63041 & 4.155(19) & 2.26(8) & 3.58(11) & 20 & 215 & 0.23 \\
NGC~2516 & $-$0.23 & HD~65712 & 4.002(19) & 1.42(12) & 2.16(9) & 90 & 870 & 0.12 \\
		&  		   & HD~65987 & 4.107(19) & 2.31(16) & [3.18] & [105] & [295] & [0.36]$^{1}$ \\
        &          & HD~66295 & 4.039(19) & 1.64(12) & 2.44(10) & 95 & 615 & 0.17 \\
		&		   & HD~66318 & 3.979(15) & 1.29(81) & 2.00(5) & 140 & 1100 & 0.10 \\
		&		   & CP$-$60\,944A & 4.104(19) & 2.10(12) & 3.14(13) & 105 & 305 & 0.33 \\
		&		   & CP$-$60\,944B & 4.096(19) & 1.96(12) & 3.03(12) & 105 & 340 & 0.32 \\
		&		   & CP$-$60\,978 & 4.076(19) & 1.90(12) & 2.82(7) & 110 & 410 & 0.26 \\
		&		   & CP$-$60\,981 & 3.975(15) & 1.51(8) & [1.95] & [100] & [1180] & [0.10]$^{2}$ \\
IC~2391 & $-$0.04 & HD~74169 & 3.998(19) & 1.41(7) & 2.25(10) & $<$\,10 & 775 & 0.07 \\
        &         & HD~74535 & 4.418(19) & 2.33(7) & 3.80(11) & 35 & 185 & 0.29 \\
IC~2602 & $-$0.20 & HD~92385 & 4.043(19) & 1.69(6) & 2.47(6) & 95 & 590 & 0.05 \\
        &         & HD~92664 & 4.176(19) & 2.47(6) & 3.99(12) & 45 & 165 & 0.18 \\
\hline
\vspace{-3mm}
\\
\multicolumn{8}{l}{$^{1}$\dots probable binary; $^{2}$\dots binary}
\end{tabular}
\end{center}
\end{table*}

\section{Determination of the evolutionary status for the CP2 stars}

For the further analysis of the evolutionary status of the CP2 objects, the program stars 
were located in a log\,$L/L_{\sun}$ versus log\,$T_{\rm eff}$ diagram to derive masses
and ages. In the following subsections we will discuss this procedure in more detail
(see also P\"ohnl 2001).

\subsection{Effective temperature}

Since all of our program stars were measured within the Geneva 7-color photometric
system (Rufener 1988) we have taken the relevant calibrations within this system.
The most recent calibration for the effective temperature in the range
from 5800 to 15000\,K is given by K\"unzli et al. (1997). However, this calibration
has to be treated with caution for CP2 stars because these objects show a ``blueing''
effect which manifests in hotter calibrated temperatures due to stronger UV absorption
than in normal type stars (Adelman 1980, Maitzen 1980). We have therefore used the
modifications for peculiar objects as given by Hauck \& North (1993) and
Hauck \& K\"unzli (1996). For the calibration within the temperatures an a-priori knowledge
of the metallicity and reddening is necessary (although the Geneva $X$ and $Y$ indices
are unaffected by reddening). The small reddening (Table\,\ref{all_res}) of our program 
clusters has been accounted for. The metallicities 
listed in the literature for IC~2602, NGC~2451\,A and NGC~2516 were taken into account
since they deviate significantly from solar values.

Hubrig et al. (2000) have compared the results of the effective temperature calibration
from the Geneva 7-color photometric system with those of spectroscopic investigations as
well as the infrared flux method and find a satisfying agreement within the range between
7500 and 15000\,K.

The accuracy of the photometric calibration for normal type stars is given between
$\pm$62 and $\pm$386\,K depending on the temperature range (Hauck \& K\"unzli 1996).
Such error values are statistically derived for a large sample of objects and have
to be treated as such. North (1998) investigated 46 CP2 stars and found an error
of the mean of 4.4\% and 3.4\% for objects hotter and cooler than 9500\,K, respectively.
These percentages transform into $\Delta$\,log\,$T_{\rm eff}$\,=\,0.019 and 0.014 which
seems reasonable also for our error estimation.

\subsection{Luminosity}

The luminosity calibration for CP2 stars is also affected by stronger
UV absorption. We have followed the approaches presented by Lanz (1984), Stepien (1994)
and North (1998) who list correction factors $\delta_{BC}$ for the estimation
of bolometric magnitudes for chemically peculiar stars. The values for $\delta_{BC}$
reach up to 0.3\,mag for objects in the range 
$-$0.85\,$<$\,$(B2-G)_0$\,$<$\,$-$0.67 within the Geneva 7-color photometric system.
The luminosity can be calculated as:
$$\log L/L_{\sun} = 0.4\cdot (4.72 -M_V -BC + \delta_{BC})$$
with 
$$M_V = m_V + 5\cdot \log \pi +5 - 3.1\cdot E(B-V).$$
The values for the bolometric correction $BC$ and the standard solar value
were taken from Flower (1996). 

We have also corrected for the so-called ``Lutz-Kelker effect''.
Lutz \& Kelker (1973) were among the first to calculate corrections
for the bias in the absolute magnitude of a star as estimated from
its trigonometric parallax. The bias is introduced by ubiquitous random
errors of measurements which, on average, cause the trigonometric
parallax to be overestimated. The correction for our program clusters range
between 0.004 and 0.028\,mag.

The error of the luminosity based on the relative error of the parallax is
given as:
$$\Delta \log L/L_{\sun} = 2\cdot\log\left(1 + \frac{\sigma(\pi)}{\pi}\right).$$
The errors of our calibrations (Table\,\ref{cp2_res}) are compatible with those
listed by e.g. North (1998).

\subsection{Age and mass}

As a next step, we have used the main sequence evolutionary tracks and isochrones
from Schaller et al. (1992), Charbonnel et al. (1993) and Schaerer et al. (1993a,b)
in order to estimate the individual masses and ages for our program stars. 
We have compared these models to those given by 
Claret \& Gimenez (1992) and Bressan (1993) and found no significant deviations 
between them (see also Lastennet et al. 1999 for such a comparison).

The models used are available for different metallicities and masses. For the 
determination
of the masses and ages we have used the grids as tabulated with the hydrogen concentration
in the core $X_{\rm C}$, log\,$L/L_{\sun}$ and log\,$T_{\rm eff}$.

Our analysis shows that, for example, the values of log\,$L/L_{\sun}$
at the zero age main sequence are nearly linear correlated with the values for
log\,$M/M_{\sun}$. The same correlation exists for the values at the
terminal age main sequence. In addition we find that the log\,$L/L_{\sun}$ and
log\,$M/M_{\sun}$ values are, at any stage of the stellar evolution, well 
correlated with $X_{\rm C}$. Since the data of the used model show that there is a 
clear dependance of the time for a star on the main sequence $\tau_{\rm HR}$ and 
log\,$M/M_{\sun}$ which can be written as
$$\log \tau_{\rm HR} = 9.99 - 3.36\cdot \log (M/M_{\sun}) +0.67\cdot\log^2(M/M_{\sun})$$
we can use the result of this quadratic regression for the relative age
$\tau/\tau_{\rm HR}$. This parameter is independent of the stellar masses and can be
described via $X_{\rm C}$ as
$$\tau/\tau_{\rm HR} = 1.004 - 0.331\cdot X_{\rm C} - 1.653\cdot X^2_{\rm C}.$$
The error for the relative age is less than $\pm$5\,\%. Finally, we have calculated
the star's ages via the above given equations. The grid between the 
zero and terminal age main sequence is defined by lines of equivalent
$M$ and $X_{\rm C}$. With known values for log\,$L/L_{\sun}$ and log\,$T_{\rm eff}$, a
box in which we have interpolated can be determined. This quadratic interpolation
results in $M$ as well as $X_{\rm C}$ and thus the age of the individual program stars.

\section{Results}

All results for the CP2 stars in the four open clusters are summarized in 
Table\,\ref{cp2_res}. In the following we will discuss the objects in the
open clusters in more detail. 

{\it IC~2391, IC~2602} and {\it NGC~2451\,A:} All CP2 stars are members
of the individual clusters with very young ages. 

{\it NGC~2516:} Figure\,\ref{ngc2516} shows the location of the CP2 stars 
together with the isochrone for log\,$t$\,=\,8.00 and $[$Fe/H$]$\,=\,$-$0.23 and the
evolutionary tracks for individual masses. All but two (HD~65987 and CP$-$60\,981) lie
very well on the apparent main sequence of this cluster and are members of it.
The two deviating objects are both binaries. Their location above the corresponding
isochrone can be very well reproduced taking the binarity into account. The masses
and ages for the visual binary system CP$-$60\,944A\,+\,B (both in very good agreement
with the values from Debernardi \& North 2001) infer that they are physically coupled. 
The mean value of the age for all six single CP2 stars is 106(17)\,Myr which is in excellent
agreement with the overall age of this cluster (100\,Myr).

The location of all investigated CP2 stars within the relevant 
Hertzsprung-Russell-diagrams qualify them as being members of the corresponding
aggregate. Furthermore, their ages (10\,$<$\,ages\,$<$\,140\,Myr)
are within the expected error of the overall age of the individual open cluster.
This proves that CP2 stars do exist at very young evolutionary stages, clearly
before they have reached 30\,\% of their main sequence life-time.

\section{Conclusions}

Several publications during the last decades have been dedicated to the 
evolutionary status of the CP2 stars. Still this issue is a matter of debate.
In principle, there are two working hypotheses: 1) CP2 stars show their
peculiar nature soon after arriving at the zero age main sequence or 2)
after about 30\,\% of their life-time on the main sequence, the
phenomenon occurs.

We have investigated four young (age not more than 100\,Myr) open clusters with
known CP2 members. The peculiarity of these objects was established via
photometric as well as spectroscopic data. The program clusters were
chosen on the basis of available Geneva 7-color photometry and known accurate
Hipparcos distances (error\,$<$\,10\,\%).

We have derived effective temperatures and luminosities for these objects
and calibrated ages as well as masses with the help of standard evolutionary
models taking the overall metallicities of the individual clusters into account.

All investigated objects are members of their open cluster with ages
between 10 and 140\,Myr corresponding to a relative age of 0.05 and
0.36 of their main sequence life-time. This clearly proves that the 
observable CP2
phenomenon occurs already well before a star has reached 30\,\% of its
life-time on the main sequence.

The somewhat discrepant result of Hubrig et al. (2000) can be understood 
by their bias in selecting objects (not by differences in the reduction
procedure): the detection
of resolved Zeeman patterns requires a specifically slow rotation. This gives 
preference to finding such objects in advanced phases on the main sequence band 
where rotational velocities have been decreased by the growth of stellar radii.

Concerning the question of a parallel evolution of both the magnetic and spectroscopic 
peculiarities, it seems still premature to draw conclusions from the existing data. A 
concentrated effort to observe both essential features of the magnetic peculiar 
stars is highly desirable in order to reveal their nature.

\begin{acknowledgements}
Use was made of the SIMBAD database, operated at CDS, Strasbourg, France.
EP acknowledges partial support from the Fonds zur F\"orderung 
der wissenschaftlichen Forschung, project {\em P14984}.
\end{acknowledgements}

\end{document}